\documentclass[twocolumn,aps,prb,superscriptaddress,amsmath,amssymb]{revtex4-2}
\usepackage{amssymb}
\usepackage{amsmath}
\usepackage{amsfonts}
\usepackage{graphicx}
\usepackage{color}
\usepackage{bm}
\usepackage[colorlinks,linkcolor=blue,anchorcolor=blue,citecolor=blue,urlcolor=blue]{hyperref}

\begin{document}

\title{Quantum anomalous Hall effect in chiral semimetals}

\author{Peng-Yi Liu}
\affiliation{International Center for Quantum Materials, School of Physics, Peking University, Beijing 100871, China}

\author{Yu-Hao Wan}
\email[Contact author: ]{wanyh@stu.pku.edu.cn}
\affiliation{International Center for Quantum Materials, School of Physics, Peking University, Beijing 100871, China}

\author{Qing-Feng Sun}
\email[Contact author: ]{sunqf@pku.edu.cn}
\affiliation{International Center for Quantum Materials, School of Physics, Peking University, Beijing 100871, China}
\affiliation{Hefei National Laboratory, Hefei 230088, China}

\begin{abstract}
The quantum anomalous Hall (QAH) effect is conventionally understood to exist only in Chern insulators, while a recent study has shown that ferromagnetic metals can also host the QAH effect.
Between insulators and metals, we demonstrate that QAH can persist even in a chiral semimetal, where conduction and valence bands touch at zero energy.
Transport calculations demonstrate that the Hall conductivity of such a system can be quantized in the presence of dephasing.
Interestingly, its longitudinal conductivity remains finite and exhibits semimetallic behavior, in contrast to Chern insulators.
This unusual transport behavior originates from the quantization of the Berry curvature integral over occupied states and the semimetallic band structure.
This chiral semimetal can transition into a Chern insulator, accompanied by the vanishing of longitudinal conductivity and a reduction of the intrinsic length scale of the Hall response.
Our results extend the concept of QAH and uncover the semimetallic QAH transport signatures.
\end{abstract}

\maketitle

\section{Introduction}
One of the most fundamental manifestations of topological physics is the quantization of the Hall conductivity.
The Hall conductivity is directly related to the integral of the Berry curvature over the occupied states, which becomes quantized in insulating systems and is characterized by an integer Chern number $\mathbb C$ \cite{Quantized_1982_Thouless,Quantal_1984_Berry,Anomalous_2010_Nagaosa}.
Such topological insulators host chiral, gapless edge states whose number is determined by the Chern number \cite{Chern_1993_Hatsugai}.
The coexistence of an insulating bulk and chiral edge states leads experimentally to a quantized Hall conductivity accompanied by a vanishing longitudinal conductivity \cite{New_1980_Klitzing,Experimental_2013_Chang}.
This phenomenon was first observed in two-dimensional electron gases under strong magnetic fields as the quantum Hall effect \cite{New_1980_Klitzing}, and was later generalized to systems without external magnetic fields, giving rise to the quantum anomalous Hall (QAH) effect in magnetic insulators \cite{Quantum_2020_Deng,Model_1988_Haldane,Topological_2006_Qi,Quantized_2010_Yu,Experimental_2013_Chang,Quantum_2010_Qiao,Tuning_2020_Zhao,Quantum_2014_Zhang,Tunable_2020_Chen}, which offers enhanced experimental accessibility and technological potential for achieving topological electronics \cite{Quantum_2023_Chang,Magnetic_2019_Tokura,Building_2021_Wu}.

Despite these advances, realizing the QAH effect under practical conditions remains challenging. Experimentally demonstrated QAH systems remain scarce, typically exhibiting small bulk gaps and low Curie temperatures that require fine-tuning and careful-designing to achieve a quantized Hall plateau \cite{Quantum_2023_Chang,Enhanced_2023_Yi}.
Identifying material platforms that can sustain robust magnetic order while supporting quantized Hall transport therefore remains a central objective in the field.
From a materials perspective, ferromagnetic materials without a bulk gap often exhibit significantly higher Curie temperatures than magnetic insulators \cite{Magnetism_2009_Coey,Anomalous_2010_Nagaosa}, suggesting a potentially broader platform for realizing QAH physics.

Recently, it has been proposed that the QAH effect can also emerge in ferromagnetic metals \cite{Quantum_2025_Wan}.
In such systems, topological chiral edge channels coexist with isotropic metallic bulk conduction.
In the presence of dephasing, the system simultaneously exhibits a disorder-robust, quantized Hall conductivity and a finite longitudinal conductivity.
This development extends the candidate materials for QAH physics from rare ferromagnetic insulators to a much broader class of ferromagnetic metals \cite{Anomalous_2010_Nagaosa,Magnetism_2009_Coey}, significantly expanding the scope for experimental realization and applications of QAH phenomena.

Between insulators and metals, there exists an important class of systems known as semimetals, in which the conduction and valence bands touch at the Fermi level.
Compared to insulators and conventional metals, semimetals exhibit remarkably rich physical phenomena and superior electronic properties \cite{Topological_2011_Wan,Weyl_2018_Armitage,The_2024_Zhai}, as exemplified by graphene \cite{Electric_2004_Novoselov,Two_2005_Novoselov,The_2009_Neto}.
The realization of QAH physics in semimetals is therefore of fundamental interest and practical importance.

In this work, we propose a minimal model of a chiral semimetal that supports a quantized anomalous Hall response.
Using nonequilibrium Green's function calculations for a six-terminal device geometry, we demonstrate that, under dephasing, the Hall conductivity is quantized, whereas the longitudinal conductivity increases away from the Fermi energy, exhibiting characteristic semimetallic behavior.
This coexistence distinguishes the chiral semimetallic QAH phase from conventional Chern insulators, where longitudinal transport is fully suppressed.
By continuously introducing a mass term, the system evolves into an insulating QAH phase, accompanied by the vanishing of the longitudinal conductivity and a notable change in the spatial profile of the equilibrium Hall current.

The rest of the paper is organized as follows.
In Sec. \ref{II}, we introduce the Hamiltonian of the chiral semimetal hosting the QAH.
In Sec. \ref{III}, we present quantum transport simulations demonstrating the emergence of QAH in the semimetal under dephasing, and we elucidate the mechanism underlying the quantization in Sec. \ref{IV}.
In Sec. \ref{V}, we compare the semimetallic QAH phase with the conventional insulating QAH phase and show the transition from the semimetallic to the insulating regime.
Finally, Sec. \ref{VI} provides discussions and conclusions.

\section{Model Hamiltonian}\label{II}
The chiral semimetal is modeled by a two-band Hamiltonian \cite{Classification_2025_Wan}:
\begin{equation}
\begin{aligned}
	H_{\rm semi}&=\sum_{x,y} \left[ c_{x,y}^\dagger T_x c_{x+1,y} +c_{x,y}^\dagger T_y c_{x,y+1} + {\rm H.c.}\right]\\
	&+\sum_{x,y}\left[ c_{x,y}^\dagger T_2 c_{x+1,y+1} + c_{x,y}^\dagger T_2 c_{x+1,y-1} + {\rm H.c.}\right],
\end{aligned}
\end{equation}
where $c_{x,y}$ and $c_{x,y}^\dagger$ are the annihilation and creation operators of electrons in a two-dimensional square lattice, respectively.
$(x,y)$ represents the site coordinate, and the lattice constant is set to the unit $a=1$.
$T_x=\frac{B}{2}\sigma_z$, $T_y=\frac{B}{2}\sigma_z+\frac{A}{2{\rm i}}\sigma_y$, and $T_2=\frac{A}{4{\rm i}}\sigma_x$ respectively represent the $x$-direction, $y$-direction, and next-nearest-neighboring hopping terms, with $2\times 2$ Pauli matrices $\sigma_{x,y,z}$.
This lattice Hamiltonian can be transformed to momentum space $H_{\rm semi}=\sum_{\bm k} c_{\bm k}^\dagger \mathcal{H}(\bm k) c_{\bm k}$, with ${\bm k}=(k_x,k_y)$.
\begin{equation}
	\begin{aligned}
	\mathcal{H}(\bm k)&= A(\sin k_x \cos k_y \sigma_x + \sin k_y \sigma_y)\\ &+ B (\cos k_x+\cos k_y) \sigma_z.
	\end{aligned}\label{eq:hk}
\end{equation}

\begin{figure}
	\centering
	\includegraphics[width=\columnwidth]{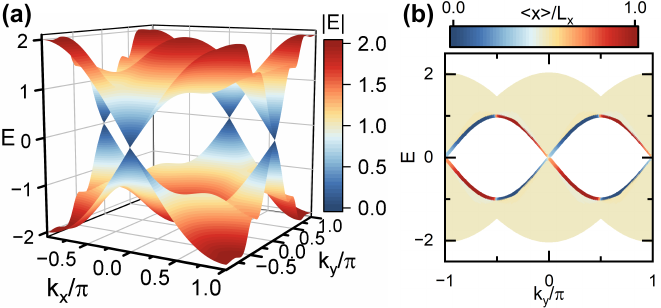}
	\caption{(a) Two-dimensional band structure of the chiral semimetal model.
	(b) Band structure of a chiral semimetal nanoribbon, with the color encoding the position expectation of each state $\langle x \rangle /L_x$
	This nanoribbon has a width in the $x$-direction $L_x=100$ and is of infinite extent in the $y$-direction.
	Throughout this paper, we use $A = B = 1$.
	}
	\label{fig:1}
\end{figure}

The band structure $E(\bm k)$ of this Hamiltonian can be obtained by diagonalizing Eq. (\ref{eq:hk}).
As shown in Fig. \ref{fig:1}(a), the band structure is symmetric about $E=0$, originating from an electron-hole symmetry of the form $\{\sigma_y \mathcal{K},\mathcal{H}(\bm k)\}=0$, where $\mathcal{K}$ represent the complex conjugate.
Notably, the electron-type and hole-type bands touch at the $X(\pi,0)$ and $Y(0,\pi)$ points, highlighting the semimetallic nature of this model.

To clearly observe the chiral characteristics of this semimetallic model, we shift our perspective to its nanoribbon band structure.
We consider a nanoribbon with an open boundary condition in the $x$-direction and infinite extent in the $y$-direction.
As shown in Fig. \ref{fig:1}(b), the band shows similar characteristics to Fig. \ref{fig:1}(a), where gapless points emerge at the center and boundary of the one-dimensional Brillouin zone.
Moreover, we use color mapping to represent the position expectation of the wave function of each Bloch state along the $x$-direction $\langle x\rangle/L_x$.
It is clear that all the high-energy states are extendedly distributed in the bulk of the nanoribbon with $\langle x\rangle/L_x\approx 0.5$.
However, for each momentum $k_y$, the wave functions of the states closest to zero energy show an obvious boundary distribution.
States with smaller $\langle x\rangle$ (shown in blue) have positive group velocities, while those with larger $\langle x\rangle$ (shown in red) have negative group velocities, which show the chiral nature of the semimetal, suggesting the potential QAH effect.

\section{Semimetallic QAH with dephasing}\label{III}
\begin{figure}
	\centerline{\includegraphics[width=\columnwidth]{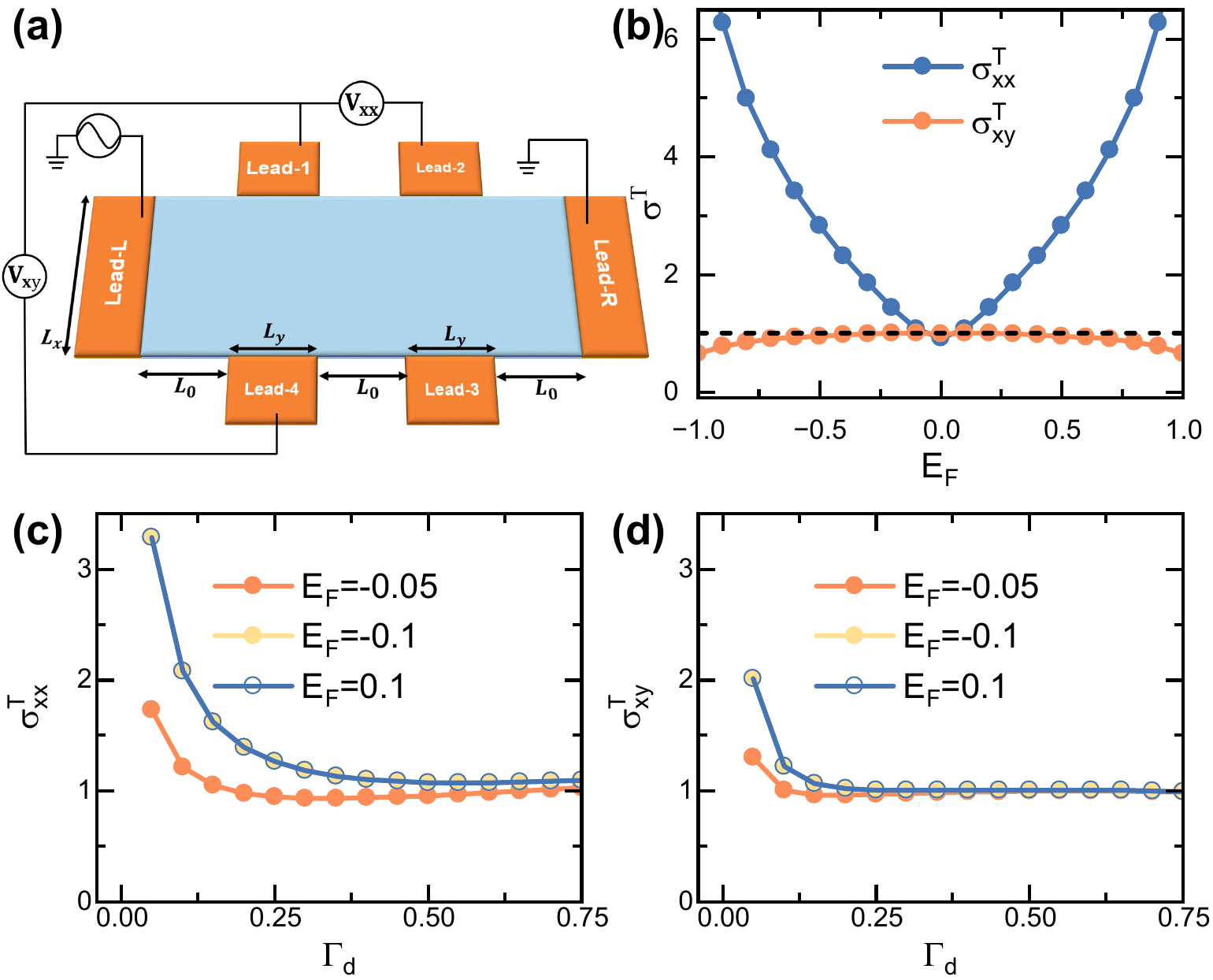}}
	\caption{(a) Schematic of a six-terminal Hall bar. Six leads are shown in orange, and the center region (shown in blue) is made of chiral semimetals.
	The width of lead-L and R is $L_x=100a$, the width of Lead-(1-4) is $L_y=50a$, and the spacing between the leads is $L_0=50a$.
	(b) Longitudinal $\sigma_{xx}^T$ and Hall $\sigma_{xy}^T$ conductivity vs. Fermi energy $E_F$, in the unit of $e^2/h$, calculated by the transport simulation. We use a dephasing strength $\Gamma_d=0.5$.
	The dashed line shows the quantum conductivity.
	(c) and (d) respectively show the $\sigma_{xx}^T$ and $\sigma_{xy}^T$ vs. $\Gamma_d$, in the unit of $e^2/h$, with three different $E_F$.
	}
	\label{fig:2}
\end{figure}

To explicitly demonstrate the quantized Hall response, we consider a realistic six-terminal Hall bar geometry and perform a standard transport simulation using nonequilibrium Green's functions.
As shown in Fig. \ref{fig:2}(a), Lead-L and Lead-R are the source and drain of the Hall bar, respectively.
Leads-(1-4) act as four voltage probes.
Furthermore, since the size of real devices significantly exceeds the micron scale \cite{Experimental_2013_Chang} and notably exceeds the typical phase-coherence length \cite{Direct_2008_Roulleau,Probing_2022_Deng}, it is necessary to introduce dephasing in the center region (shown in blue) to simulate the realistic transport behavior.
We use the B\"{u}ttiker virtual leads to introduce the dephasing mechanism \cite{Four_1986_Buttiker,Quantum_2025_Wan,Dissipation_2024_Liu}, so the electrons can enter the virtual leads and then flow back into the system, accompanied by the loss of phase coherence but without net particle loss.
In our simulations, half of the sites in the center region are coupled to virtual leads, which are labelled as Leads-(5,6,7,...).
The current of lead-$p$ ($I_p$) is determined by the Landauer-B\"{u}ttiker formula \cite{Landauer_1992_Meir,Electronic_1995_Datta}:
\begin{equation}
I_p = \sum_q T_{pq} (V_p-V_q),\label{eq:landauer}
\end{equation}
where $V_p$ is the voltage of Lead-$p$, with $p$ and $q$ running over both real and virtual leads ($p,q=L,R,1,2,3,...$).
The transmission coefficient is given by $T_{pq}(E_F)={\rm Tr} \left[{\bf \Gamma}_p(E_F) {\bf G}^R_{pq}(E_F) {\bf \Gamma}_q(E_F) {\bf G}^A_{qp}(E_F)\right]$, where ${\bf G}^R=({\bf G}^A)^\dagger$ is the retarded Green's function.
Using the Dyson's equation, ${\bf G}^R(E_F)=[(E_F+{\rm i}0^{+})-{\bf H}_{\rm cen}-\sum_p {\bf \Sigma}_p^{R}(E_F)]^{-1}$, where ${\bf H}_{\rm cen}$ is the Hamiltonian of the center region.
${\bf \Sigma}_p^{R}$ is the retarded self-energy coming from the coupling of Lead-$p$, and the corresponding linewidth is defined as ${\bf \Gamma}_p={\rm i}\left[{\bf \Sigma}_p^{R} -({\bf \Sigma}_p^{R})^\dagger \right]$.
For real leads ($p=L,R,1,2,3,4$), we use ${\bf \Sigma}_p^{R}=-{\rm i}{\bf I_p}/2$, and for virtual leads ($p=5,6,7,...$), ${\bf \Sigma}_p^{R}=-{\rm i} \Gamma_d{\bf I_p}/2$, where ${\bf I_p}$ is the identity matrix of sites coupled to Lead-$p$ and $\Gamma_d$ is the dephasing strength.
In order to simulate a real experiment, we use a small bias voltage between the source and drain: $V_L=V$ and $V_R=0$.
Since no current leaves the system during the voltage measurement and dephasing process, we set $I_{1,2,3...}=0$.
With Eq. (\ref{eq:landauer}) and these boundary conditions, we can obtain the current $I_L=-I_R$ and all the voltages $V_{1,2,3,...}$.

To explore the longitudinal and Hall conductivity under the transport simulation, we first measure the longitudinal and Hall resistivity: $\rho_{xx}=\frac{(V_{1}-V_{2})}{I_L}\frac{L_x}{L_0+L_y}$ and $\rho_{xy}=(V_{1}-V_{4})/I_L$.
Then, the conductivities are obtained by an inversion of the resistivity tensor \cite{Quantum_2025_Wan,Transport_2022_Zhou}:
\begin{equation}
	\sigma_{xx}^T=\frac{\rho_{xx}}{\rho_{xx}^2+\rho_{xy}^2},\quad\sigma_{xy}^T=\frac{\rho_{xy}}{\rho_{xx}^2+\rho_{xy}^2},\label{eq:inverse}
\end{equation}
where the superscript ``T'' means that the results come from transport simulations.
The results with an appropriate dephasing strength $\Gamma_d$ are shown in Fig. \ref{fig:2}(b).
When $E_F=0$, the longitudinal conductivity is small due to the low density of states at $E=0$ of the semimetal.
However, as $|E_F|$ increases, $\sigma_{xx}^T$ increases almost linearly, which is a typical characteristic of semimetals \cite{The_2009_Neto}, different from insulators and metals.
Importantly, around $E_F=0$, the Hall conductivity is quantized to the quantum conductivity $e^2/h$, as a key signature of QAH.
This wide energy window of quantized Hall conductivity demonstrates that the QAH effect can be robustly realized in a semimetallic system despite the presence of bulk carriers.
In contrast to the QAH phenomenon in traditional insulators, a finite longitudinal conductivity and quantized Hall conductivity occur simultaneously.
At this time, the longitudinal resistance is not zero, and the Hall resistance is not quantized even though the Hall conductivity remains quantized, which closely resembles the recently proposed QAH effect in ferromagnetic metals \cite{Quantum_2025_Wan}.

Next, we vary the dephasing strength $\Gamma_d$ to examine the conditions under which QAH occurs in semimetals.
As shown in Fig. \ref{fig:2}(c), regardless of how $\Gamma_d$ changes, the longitudinal conductivity always remains finite, marking a distinct difference from the conventional QAH phenomenon found in insulators.
More interestingly, as shown in Fig. \ref{fig:2}(d), when $\Gamma_d$ is small, the Hall conductivity deviates from the quantum conductivity.
In contrast, when the dephasing is strong enough, the Hall conductivity converges to the quantized plateau for different Fermi energies.
This result reveals an unconventional phenomenon: although the QAH originates from quantum effects, its experimental manifestation in semimetals/metals requires partial suppression of long-range quantum coherence \cite{Experimental_2022_Mogi,Transport_2022_Zhou,Quantum_2025_Wan,Dephasing_2023_Fang}.
In fully phase-coherent quantum transport, the electronic states cannot be characterized by a spatially local and uniform resistivity tensor, such as in the case of ballistic transport \cite{Electronic_1995_Datta,Voltage_1991_McLennan}.
Therefore, Eq.~({\ref{eq:inverse}}) is breakdown and $\sigma_{xy}^T$ deviates from the ideal value.
Fortunately, in realistic devices, dephasing is inevitable \cite{Weak_1986_Chakravarty,Phase_1990_Stern,Coupled_1999_Burkard}, and the size of Hall bars is typically much larger than the phase-coherence length \cite{Experimental_2013_Chang,Direct_2008_Roulleau,Probing_2022_Deng}, pulling the perfect quantum transport back to classical.
In this regime, the quantized Hall conductivity can be reliably extracted using standard six-terminal measurements \cite{Transport_2022_Zhou,Quantum_2025_Wan,Dephasing_2023_Fang}, as demonstrated in Fig.~\ref{fig:2}(d).

As an experimental perspective, Figs.~\ref{fig:2}(b) and \ref{fig:2}(d) illustrate possible routes for observing the Hall plateau in a semimetallic system.
In experiments on two-dimensional materials, the carrier density or Fermi energy can typically be tuned by a gate voltage, which may allow one to observe a Hall-conductivity plateau as a function of gate voltage \cite{Experimental_2013_Chang}, similar to the behavior shown in Fig. \ref{fig:2}(b).
In addition, the dephasing strength in the system is strongly influenced by temperature, as the phase-coherence length $\propto T^{-p/2}$ in low temperatures, where $T$ is the temperature and $p\approx2$ \cite{Probing_2022_Deng}.
Therefore, it is expected that tuning the temperature may enable the observation of a Hall plateau as a function of temperature \cite{Experimental_2022_Mogi}, analogous to the behavior with dephasing shown in Fig. \ref{fig:2}(d).
Furthermore, as discussed in Appendix \ref{A}, we have systematically examined the size dependence of the transport coefficients.
The Hall conductivity is found to be robust against variations in the device dimensions, while the longitudinal conductivity shows noticeable size dependence.
These results may provide useful guidance for experimental realization.

\section{Origin of the quantized Hall response}\label{IV}
To elucidate the microscopic origin of the quantized Hall response in chiral semimetals, we revisit the structure of the lattice Hamiltonian in Eq.~(\ref{eq:hk}).
The first term generates four Dirac cones required by fermion doubling, located at $\Gamma(0,0)$, $X(\pi,0)$, $Y(0,\pi)$, and $M(\pi,\pi)$, while the second term introduces momentum-dependent mass terms for these cones.
Expanding the Bloch Hamiltonian around $\Gamma$ and $M$, $\bm k=\bm k_{\Gamma,M}+\bm q$, we obtain
$\mathcal{H}_{\Gamma,M}(\bm q)=A(q_x\sigma_x \pm q_y\sigma_y)\pm 2B\sigma_z$,
which describes two fully gapped Dirac fermions with opposite winding numbers and opposite masses.
Because each massive Dirac cone contributes a half-quantized Berry curvature integral, their combination yields an integer and uncompensated contribution to the Hall response \cite{Topological_2013_Bernevig}.

Because Berry curvature is typically concentrated near band-touching points, the presence of gapless cones at $X$ and $Y$ naturally raises a question: Whether their contributions could interfere with the quantization of the Hall response?
The low-energy Hamiltonian around these two points takes the form
\begin{equation}
\mathcal{H}_{X,Y}(\bm q)
= A(-q_x\sigma_x \pm q_y\sigma_y)
\pm \frac{B}{2}(q_x^2-q_y^2)\sigma_z,\label{eq:lowenergy}
\end{equation}
where the upper (lower) sign corresponds to the $X$ ($Y$) point.
Equation (\ref{eq:lowenergy}) describes two Dirac fermions with opposite winding, whose leading mass term is quadratic in momentum and exhibits a $d$-wave form.
This momentum-dependent mass changes sign under a $\pi/2$ rotation and has recently been identified as an altermagnetic mass term \cite{Emergent_2026_Liu,Interplay_2025_Wan,Helical_2025_Wan}.
As a consequence of this $d$-wave sign structure, the Berry curvature near the $X$ and $Y$ points displays a characteristic pattern of alternating positive and negative contributions, whose integrals vanish exactly \cite{Emergent_2026_Liu}.
This ensures that the gapless cones do not destroy the quantized Hall response originating from the $\Gamma$ and $M$ points.

\begin{figure}
	\centerline{\includegraphics[width=1\columnwidth]{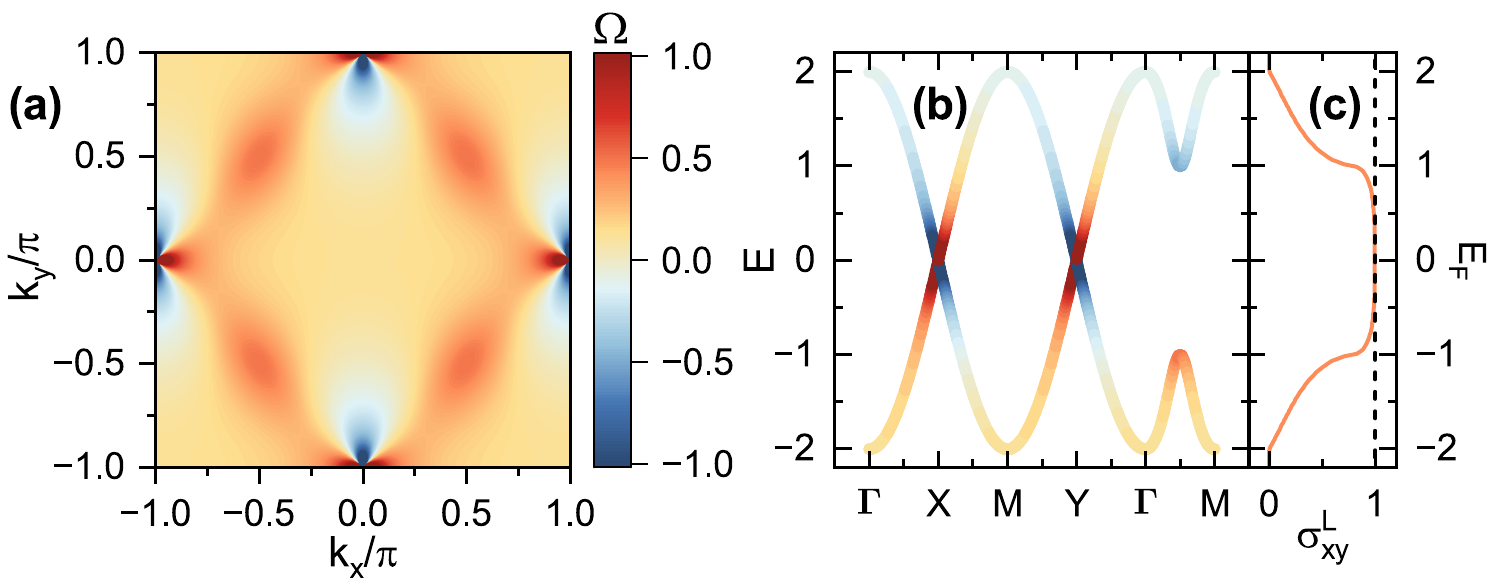}}
	\caption{
(a) Berry curvature $\Omega$ distribution of the lower band of Eq. (\ref{eq:hk}) in the Brillouin zone.
(b) Band structure plotted along high-symmetry lines, with the Berry curvature encoded by the color scale, using the same color bar as in panel (a).
(c) Linear-response Hall conductivity $\sigma_{xy}^L$, obtained by integrating the Berry curvature of all occupied states below the Fermi energy, as a function of the Fermi energy, in units of $e^2/h$.}
	\label{fig:3}
\end{figure}

This mechanism is further illustrated in Fig.~ \ref{fig:3}.
Figure~\ref{fig:3}(a) shows the Berry curvature distribution of the lower band, where strong but sign-alternating Berry curvature appears around the gapless $X$ and $Y$ points.
When mapped onto the band structure [Fig.~\ref{fig:3}(b)], these contributions cancel upon integration, while the Berry curvature between $\Gamma$ and $M$ remains finite and uncompensated.
Integrating the Berry curvature over the occupied states yields the linear-response Hall conductivity \cite{Topological_2013_Bernevig} shown in Fig.~\ref{fig:3}(c), which remains quantized as the Fermi energy is varied within $-0.5 \lesssim E_F \lesssim 0.5$.
This quantized integral of the Berry curvature provides the microscopic origin of the Hall plateaus observed in the transport calculations of Fig.~\ref{fig:2}.

\section{Insulating QAH and semimetal-insulator transition}\label{V}
\begin{figure}
	\centerline{\includegraphics[width=1\columnwidth]{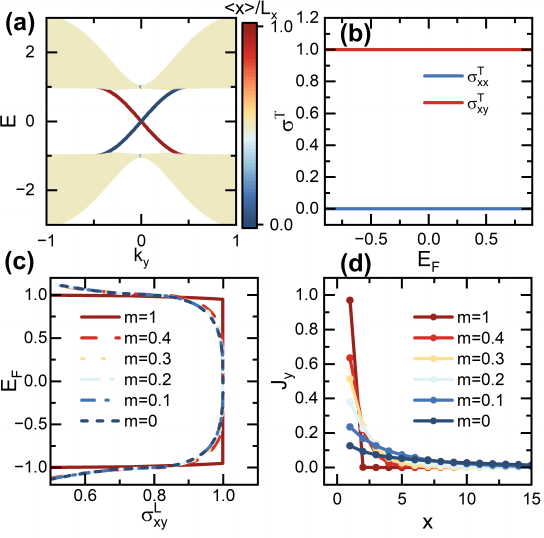}}
	\caption{
		(a) Band structure of the insulating QAH nanoribbon with the mass $m=1$.
		This nanoribbon has a width in the $x$-direction $L_x=100$ and is of infinite extent in the $y$-direction.
		(b) The longitudinal $\sigma_{xx}^T$ and Hall $\sigma_{xy}^T$ conductivities calculated by the transport simulations, with the unit of $e^2/h$ and $\Gamma_d=0$.
		In (a) and (b), $m=1$.
		(c) Linear-response Hall conductivity $\sigma_{xy}^L$ for different $m$, obtained by integrating the Berry curvature of all occupied states below the Fermi energy $E_F$, as a function of the Fermi energy, in units of $e^2/h$.
		(d) Distribution of the $y$-direction equilibrium current along the $x$ position of a nanoribbon, whose width in the $x$-direction is $L_x=50$.	Here, we use $E=0$ and $\eta=0.03$.
		}
	\label{fig:4}
\end{figure}

As a contrast to the semimetallic QAH phase discussed above, the same lattice model can be continuously tuned into a conventional QAH insulator by introducing a uniform mass term: $H_{\rm insu}=H_{\rm semi}+\sum_{x,y} m c_{x,y}^\dagger \sigma_z c_{x,y}$ with the mass $m$.
This term completes the opening of a full bulk gap by gapping the previously gapless cones at the $X$ and $Y$ points.
The nanoribbon band structure of the resulting QAH insulator is shown in Fig.~\ref{fig:4}(a), where a clear bulk energy gap is opened.
The color mapping shows that the bulk states are distributed in the bulk of the nanoribbon.
As dictated by the bulk-edge correspondence, chiral edge states traverse the bulk gap, with opposite propagation directions on opposite edges of the nanoribbon, characteristic of a conventional QAH insulator \cite{Topological_2006_Qi}.
We perform the same transport simulation as the semimetallic case for this insulating case, and the results are presented in Fig. \ref{fig:4}(b).
It is clear that the Hall conductivity is quantized within the band gap, just as expected.
But the longitudinal conductivity remains zero, which is the nature of QAH insulators \cite{Experimental_2013_Chang,Quantum_2023_Chang}.
Owing to the absence of longitudinal conduction, the Hall conductivity can be directly extracted from $I_L/(V_1-V_4)$ and remains quantized when $\Gamma_d=0$ without the requirement of a well-defined local resistivity.
This behavior contrasts with the semimetallic case and reflects a fundamental difference between insulating and semimetallic realizations of the QAH effect.

Although both the semimetallic and insulating regimes exhibit a quantized Hall conductivity, their microscopic mechanisms differ in an essential way.
As discussed above, the quantized Hall response in the semimetal originates from the two massive Dirac cones at the $\Gamma$ and $M$ points, while the gapless cones at $X$ and $Y$ give no net contribution due to their $d$-wave-symmetry-enforced cancellation.
Upon introducing a finite mass $m$, the $X$ and $Y$ cones are also gapped, and their low-energy Hamiltonian becomes $\mathcal{H}_{X,Y}(\bm q) = A(-q_x\sigma_x \pm q_y\sigma_y) + \left[m\pm\frac{B}{2}(q_x^2-q_y^2)\right]\sigma_z$.
Importantly, the low-energy Hamiltonian satisfies the symmetry $\mathcal{H}_X(q_x,q_y)=\mathcal{U}\mathcal{H}_Y(q_y,q_x)\mathcal{U}^\dagger$, where $\mathcal{U}=\exp ({\rm i}\pi\sigma_z/4)$.
As a consequence, their Berry curvatures obey $\Omega_X(q_x,q_y)=-\mathcal{U}\Omega_Y(q_y,q_x) \mathcal{U}^\dagger$, and the integral of $X$ and $Y$ valley cancels exactly.
Therefore, the total Hall response continues to be entirely determined by the $\Gamma$ and $M$ points.

This continuity of the Hall response across the semimetal-insulator transition is further illustrated in Fig.~\ref{fig:4}(c), where we plot the linear-response Hall conductivity $\sigma_{xy}^L$ for different values of $m$.
For all cases, the Hall conductivity remains quantized as long as the Fermi energy lies in the vicinity of zero energy.
Only when $|E_F|$ exceeds both the mass scale $m$ and the validity range of the low-energy description does $\sigma_{xy}^L$ begin to deviate from its quantized value.
This behavior reflects the fact that the $X$ and $Y$ valleys contribute neither in the semimetallic limit ($m=0$) nor after they are gapped ($m\neq 0$).
Across the entire transition, their combined Berry-curvature integral remains zero, while the quantized contribution from the $\Gamma$ and $M$ points is preserved.

While the quantized Hall conductivity remains unchanged across the transition, the spatial structure of the underlying current response undergoes a transformation.
To characterize this difference, we examine the spatial attenuation of the equilibrium current of a nanoribbon along the $y$-direction.
The spatial-resolved equilibrium current flowing along the $y$-direction can be expressed as \cite{Altermagnetism_2025_Wan}:
\begin{equation}
	\begin{aligned}
J_{y}(x,E)&=-\frac{e}{h} \int_{-\pi}^{\pi}{\rm d}k_y\\
&\Im \left\{{\rm Tr} \left[ \left\langle x \left|\frac{\partial \mathcal{H}(k_y)}{\partial k_y} \frac{1}{E-\mathcal{H}(k_y)+{\rm i}\eta} \right| x \right\rangle\right]\right\}.
	\end{aligned}
\end{equation}
As shown in Fig.~\ref{fig:4}(d), the equilibrium current decays away from the edge at $x=1$.
In the semimetallic case $m=0$, the decay is slow, and the current penetrates deeply into the bulk, indicating the semimetallic and gapless band structure \cite{Half_2022_Zou,Altermagnetism_2025_Wan}.
With increasing $|m|$, the decay becomes progressively faster, and in the insulating regime, the current exhibits conventional QAH edge states, becoming almost entirely confined to the edges \cite{Topological_2006_Qi}.
These results indicate that although the Hall conductivity remains in a quantized value throughout the transition process, the existence of a bulk energy gap significantly affects the intrinsic length scale of the Hall response.

\section{Discussions and conclusions}\label{VI}
The quantum anomalous Hall effect represents a cornerstone for realizing robust topological electronic functionalities.
Extending QAH physics beyond fully gapped insulators to metallic and semimetallic systems is therefore of both fundamental and practical interest.
From a materials perspective, this extension substantially broadens the landscape of potential QAH candidates, as the presence of a semimetallic bulk band structure does not \emph{a priori} preclude a quantized Hall response.
Ferromagnetic metals and semimetals are considerably more abundant than ferromagnetic insulators and often exhibit higher Curie temperatures and more robust magnetic order \cite{Anomalous_2010_Nagaosa,Magnetism_2009_Coey}, potentially providing a wider materials platform for realizing QAH phenomena.
From an experimental viewpoint, the quantized Hall plateau response in our model persists over a relatively large Fermi-energy window, which may facilitate experimental observation since the carrier density or Fermi energy in two-dimensional systems can typically be tuned in a small range through gate voltages.
In addition, since dephasing is strongly influenced by temperature, temperature control may provide another practical route for probing the predicted Hall plateaus.

One of the key motivations for studying the QAH effect in insulating systems, although recently challenged \cite{Imaging_2019_Marguerite,Thermal_2021_Fang}, is the possibility of dissipationless transport carried by chiral edge states.
In the semimetallic QAH phase considered here, bulk carriers are present at the Fermi level, so dissipation is more unavoidable.
However, our findings carry a distinct significance:
they uncover a topological transport regime that extends beyond the conventional insulating QAH paradigm, demonstrating that a quantized transverse response can coexist with metallic longitudinal conduction.
From a conceptual perspective, this finding shows that a full bulk gap is not a necessary prerequisite for quantized anomalous Hall transport.
From a practical viewpoint, the coexistence of quantized Hall response and metallic conduction introduces new possibilities for device design beyond traditional insulating QAH \cite{Current_2017_Kawamura}.
Specifically, the Hall channel can serve as a robust and quantized transverse readout, which is ideal for sensing or logic applications.
At the same time, the longitudinal channel remains conductive and electrically accessible, offering a direct pathway for current flow and active modulation.
This dual-channel functionality, integrating both quantization and tunable conductivity within the same phase, may offer added versatility for future topological electronic devices.

In conclusion, we construct a distinct class of QAH phases that combine a quantized Berry curvature integral with a semimetallic energy spectrum.
Despite the absence of a full bulk gap, the Hall conductivity remains quantized, while the longitudinal conductivity increases with the absolute value of the Fermi energy due to semimetallic bulk carriers.
Using the nonequilibrium Green's function, we show that this coexistence can be captured within a six-terminal transport measurement.
We further demonstrate a transition from a semimetallic QAH phase to a QAH insulator, accompanied by the suppression of longitudinal conductivity and a decrease in the equilibrium current decay length.
These results establish a clear distinction between semimetallic and insulating realizations of QAH physics, while also pointing to broader possibilities for topological phases beyond the insulating paradigm.

\section*{Acknowledgments}
This work was financially supported
by the National Key R and D Program of China (Grant No. 2024YFA1409002),
the National Natural Science Foundation of China (Grants No. 12374034 and Grants No. 124B2069),
and the Quantum Science and Technology-National Science and Technology Major Project (Grant No. 2021ZD0302403).
We acknowledge the High-performance Computing Platform of Peking University
for providing computational resources.

\section*{Data availability}
The data that support the findings of this article are not publicly available. The data are available from the authors upon reasonable request.

\appendix
\section{Size dependence of the conductivities}\label{A}

\begin{figure}
	\centering
	\includegraphics[width=\columnwidth]{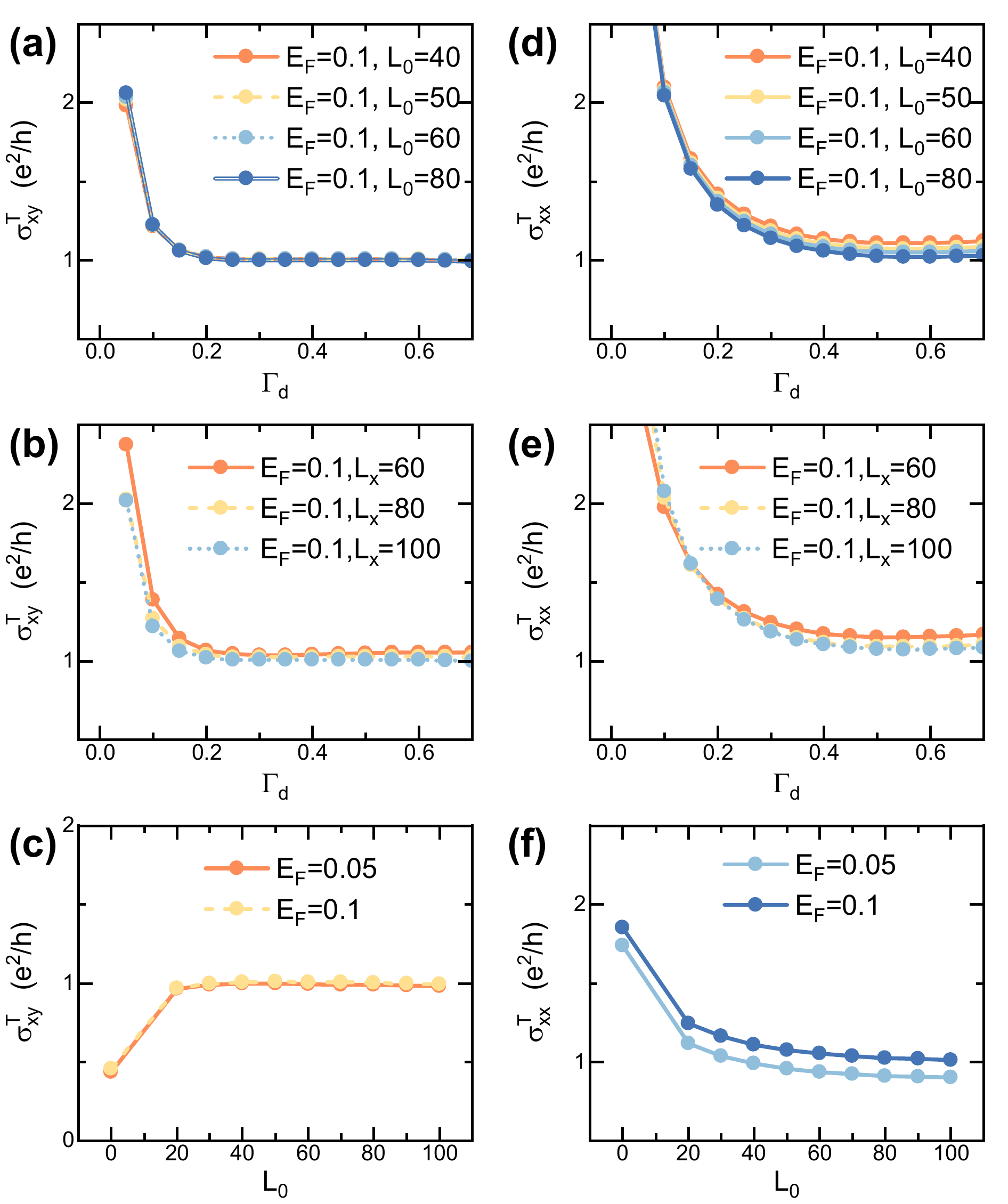}
	\caption{
	(a) the Hall conductivity $\sigma_{xy}^T$ vs. the dephasing strength $\Gamma_d$, with four different $L_0$ and $E_F=0.1$.
	(b) $\sigma_{xy}^T$ vs. $\Gamma_d$, with three different $L_x$ and $E_F=0.1$.
	(c) $\sigma_{xy}^T$ vs. $L_0$, with $\Gamma_d=0.5$ and $E_F=0.05,0.1$.
	(d), (e), and (f) the corresponding longitudinal conductivity $\sigma_{xx}^T$.
	As shown in Fig. \ref{fig:2}(a) of the manuscript, $L_x$ is the width of the Hall bar, $L_0$ is the space between voltage probes, controlling the length of the Hall bar.
	Other parameters are the same as Fig. \ref{fig:2}.}
	\label{fig:5}
\end{figure}

In this Appendix we examine the dependence of the transport coefficients in Fig. \ref{fig:2} on the device dimensions in the six-terminal Hall-bar geometry.
The results are summarized in Fig.~\ref{fig:5}.

Figures~\ref{fig:5}(a) and \ref{fig:5}(b) show the Hall conductivity $\sigma_{xy}^T$ as a function of the dephasing strength $\Gamma_d$ for several different device lengths $L_0$ and widths $L_x$, respectively.
In both cases the Hall conductivity converges to the quantized value over a broad range of $\Gamma_d$, indicating that the quantized Hall response is robust against variations of the system dimensions.
In addition, as shown in Fig.~\ref{fig:5}(c), when the distance between the voltage probes $L_0$ increases from zero, the Hall conductivity rapidly reaches the quantized plateau and remains nearly insensitive to the Fermi energy $E_F$.
These results demonstrate that the quantized Hall conductivity originates from the topological Berry-curvature integral and is therefore robust to moderate changes in the device geometry.

In contrast, the longitudinal conductivity $\sigma_{xx}^T$ shows a noticeable dependence on the system dimensions. Figures~\ref{fig:5}(d) and \ref{fig:5}(e) display $\sigma_{xx}^T$ as a function of the dephasing strength $\Gamma_d$ for different values of $L_0$ and $L_x$.
Although $\sigma_{xx}^T$ may take values close to $e^2/h$ in certain parameter ranges, its magnitude varies with both the length and width of the Hall bar.
This behavior is further illustrated in Fig.~\ref{fig:5}(f), where $\sigma_{xx}^T$ is plotted as a function of $L_0$ for several parameter sets; the longitudinal conductivity changes notably with the system size and even cross the value $e^2/h$.

These behavior reflects the different physical origins of the two conductivities.
The Hall conductivity is determined by the Berry curvature integrated over the occupied states and is therefore topologically robust.
By contrast, the longitudinal conductivity arises from bulk carriers in the semimetallic regime and depends on the details of transport, such as the precise Fermi energy and device geometry.


\begin{thebibliography}{99}

	\bibitem{Quantized_1982_Thouless}
	D. J. Thouless, M. Kohmoto, M. P. Nightingale, and M. den Nijs, Quantized Hall Conductance in a Two-Dimensional Periodic Potential, \href{https://doi.org/10.1103/PhysRevLett.49.405}{Phys. Rev. Lett. \textbf{49}, 405 (1982)}.

	\bibitem{Quantal_1984_Berry}
	M. V. Berry, Quantal phase factors accompanying adiabatic changes, \href{https://doi.org/10.1098/rspa.1984.0023}{Proc. R. Soc. London, Ser. A \textbf{392}, 45 (1984)}.

	\bibitem{Anomalous_2010_Nagaosa}
	N. Nagaosa, J. Sinova, S. Onoda, A. H. MacDonald, and N. P. Ong, Anomalous Hall effect, \href{https://doi.org/10.1103/RevModPhys.82.1539}{Rev. Mod. Phys. \textbf{82}, 1539 (2010)}.

	\bibitem{Chern_1993_Hatsugai}
	Y. Hatsugai, Chern number and edge states in the integer quantum Hall effect, \href{https://doi.org/10.1103/PhysRevLett.71.3697}{Phys. Rev. Lett. \textbf{71}, 3697 (1993)}.

	\bibitem{New_1980_Klitzing}
	K. v. Klitzing, G. Dorda, and M. Pepper, New Method for High-Accuracy Determination of the Fine-Structure Constant Based on Quantized Hall Resistance, \href{https://doi.org/10.1103/PhysRevLett.45.494}{Phys. Rev. Lett. \textbf{45}, 494 (1980)}.

	\bibitem{Experimental_2013_Chang}
	C.-Z. Chang, J. Zhang, X. Feng, J. Shen, Z. Zhang, M. Guo, K. Li, Y. Ou, P. Wei, L.-L. Wang, Z.-Q. Ji, Y. Feng, S. Ji, X. Chen, J. Jia, X. Dai, Z. Fang, S.-C. Zhang, K. He, Y. Wang, L. Lu, X.-C. Ma, and Q.-K. Xue, Experimental Observation of the Quantum Anomalous Hall Effect in a Magnetic Topological Insulator, \href{https://doi.org/10.1126/science.1234414}{Science \textbf{340}, 167-170 (2013)}.

	\bibitem{Quantum_2020_Deng}
	Y. Deng, Y. Yu, M. Z. Shi, Z. Guo, Z. Xu, J. Wang, X. H. Chen, and Y. Zhang, Quantum anomalous Hall effect in intrinsic magnetic topological insulator MnBi2Te4, \href{https://doi.org/10.1126/science.aax8156}{Science \textbf{367}, 895 (2020)}

	\bibitem{Tuning_2020_Zhao}
	Y. F. Zhao, R. X. Zhang, R. B. Mei, L. J. Zhou, H. M. Yi, Y. Q. Zhang, J. B. Yu, R. Xiao, K. Wang, N. Samarth, M. H. W. Chan, C. X. Liu, C. Z. Chang, Tuning the Chern number in quantum anomalous Hall insulators, \href{https://doi.org/10.1038/s41586-020-3020-3}{Nature \textbf{588}, 419 (2020)}.

	\bibitem{Tunable_2020_Chen}
	G. Chen, A. L. Sharpe, E. J. Fox, Y.-H. Zhang, S. Wang, L. Jiang, B. Lyu, H. Li, K. Watanabe, T. Taniguchi, Z. Shi, T. Senthil, D. Goldhaber-Gordon, Y. Zhang, and F. Wang, Tunable correlated Chern insulator and ferromagnetism in a moiré superlattice, \href{https://doi.org/10.1038/s41586-020-2049-7}{Nature \textbf{579}, 56 (2020)}.

	\bibitem{Model_1988_Haldane}
	F. D. M. Haldane, Model for a Quantum Hall Effect without Landau Levels: Condensed-Matter Realization of the ``Parity Anomaly", \href{https://doi.org/10.1103/PhysRevLett.61.2015}{Phys. Rev. Lett. \textbf{61}, 2015 (1988)}.

	\bibitem{Topological_2006_Qi}
	X.-L. Qi, Y.-S. Wu, and S.-C. Zhang, Topological quantization of the spin Hall effect in two-dimensional paramagnetic semiconductors, \href{https://doi.org/10.1103/PhysRevB.74.085308}{Phys. Rev. B \textbf{74}, 085308 (2006)}.

	\bibitem{Quantized_2010_Yu}
	R. Yu, W. Zhang, H.-J. Zhang, S.-C. Zhang, X. Dai, and Z. Fang, Quantized Anomalous Hall Effect in Magnetic Topological Insulators, \href{https://doi.org/10.1126/science.1187485}{Science \textbf{329}, 61 (2010)}.

	\bibitem{Quantum_2014_Zhang}
	H. Zhang, Y. Xu, J. Wang, K. Chang, and S.-C. Zhang,
	Quantum Spin Hall and Quantum Anomalous Hall States Realized in Junction Quantum Wells,
	\href{https://journals.aps.org/prl/abstract/10.1103/PhysRevLett.112.216803}{Phys. Rev. Lett. \textbf{112}, 216803 (2014)}.

	\bibitem{Quantum_2010_Qiao}
	Z. Qiao, S. A. Yang, W. Feng, W.-K. Tse, J. Ding, Y. Yao2, J. Wang, and Q. Niu, Quantum anomalous Hall effect in graphene from Rashba and exchange effects, \href{https://doi.org/10.1103/PhysRevB.82.161414}{Phys. Rev. B \textbf{82}, 161414(R) (2010)}.

	\bibitem{Quantum_2023_Chang}
	C.-Z. Chang, C.-X. Liu, and A. H. MacDonald, \emph{Colloquium}: Quantum anomalous Hall effect, \href{https://doi.org/10.1103/RevModPhys.95.011002}{Rev. Mod. Phys. \textbf{95}, 011002 (2023)}.

	\bibitem{Magnetic_2019_Tokura}
	Y. Tokura, K. Yasuda, A. Tsukazaki, Magnetic topological insulators, \href{https://doi.org/10.1038/s42254-018-0011-5}{Nat. Rev. Phys. 1, 126 (2019)}.

	\bibitem{Building_2021_Wu}
	B.-L. Wu, Z.-B. Wang, Z.-Q. Zhang, and H. Jiang, Building programmable integrated circuits through disordered Chern insulators, \href{https://doi.org/10.1103/PhysRevB.104.195416}{Phys. Rev. B 104, 195416 (2021)}.

	\bibitem{Enhanced_2023_Yi}
	H. T., Yi, D. Jain, X. Yao, and S. Oh, Enhanced Quantum Anomalous Hall Effect with an Active Capping Layer, \href{https://doi.org/10.1021/acs.nanolett.3c01313}{Nano Lett. \textbf{23}, 5673 (2023)}.

	\bibitem{Magnetism_2009_Coey}
	J. M. D. Coey, Magnetism and Magnetic Materials (Cambridge University Press, Cambridge, 2009).

	\bibitem{Quantum_2025_Wan}
	Y.-H. Wan, P.-Y. Liu, and Q.-F. Sun, Quantum Anomalous Hall Effect in Ferromagnetic Metals, \href{https://doi.org/10.1103/8vs2-jvc4}{Phys. Rev. Lett. \textbf{135}, 186302 (2025)}.

	\bibitem{Topological_2011_Wan}
	X. Wan, A. M. Turner, A. Vishwanath, and S. Y. Savrasov, Topological semimetal and Fermi-arc surface states in the electronic structure of pyrochlore iridates, \href{https://doi.org/10.1103/PhysRevB.83.205101}{Phys. Rev. B \textbf{83}, 205101 (2011)}.
	\bibitem{Weyl_2018_Armitage}
	N. P. Armitage, E. J. Mele, and A. Vishwanath, Weyl and Dirac semimetals in three-dimensional solids, \href{https://doi.org/10.1103/RevModPhys.90.015001}{Rev. Mod. Phys. \textbf{90}, 015001 (2018)}.
	\bibitem{The_2024_Zhai}
	E. Zhai, T. Liang, R. Liu, M. Cai, R. Li, Q. Shao, C. Su, and Y. C. Lin, The rise of semi-metal electronics, \href{https://doi.org/10.1038/s44287-024-00068-z}{Nat. Rev. Electr. Eng. \textbf{1}, 497 (2024)}.

	\bibitem{Electric_2004_Novoselov}
	K. S. Novoselov, A. K. Geim, S. V. Morozov, D. Jiang, Y. Zhang, S. V. Dubonos, I. V. Grigorieva, and A. A. Firsov, Electric Field Effect in Atomically Thin Carbon Films, \href{https://doi.org/10.1126/science.1102896}{Science \textbf{306}, 666 (2004)}.
	\bibitem{Two_2005_Novoselov}
	K. S. Novoselov, A. K. Geim, S. V. Morozov, D. Jiang, M. I. Katsnelson, I. V. Grigorieva, S. V. Dubonos, and A. A. Firsov, Two-dimensional gas of massless Dirac fermions in graphene, \href{https://www.nature.com/articles/nature04233}{Nature \textbf{438}, 197 (2005)}.
	\bibitem{The_2009_Neto}
	A. H. Castro Neto, F. Guinea, N. M. R. Peres, K. S. Novoselov, and A. K. Geim, The electronic properties of graphene, \href{https://doi.org/10.1103/RevModPhys.81.109}{Rev. Mod. Phys. \textbf{81}, 109 (2009)}.

	\bibitem{Classification_2025_Wan}
	Y.-H. Wan, P.-Y. Liu, and Q.-F. Sun, Classification of Chern numbers based on high-symmetry points, \href{https://doi.org/10.1103/PhysRevB.111.L161410}{Phys. Rev. B \textbf{111}, L161410 (2025)}.

	\bibitem{Direct_2008_Roulleau}
	P. Roulleau, F. Portier, D. C. Glattli, P. Roche, A. Cavanna, G. Faini, U. Gennser, and D. Mailly, Direct measurement of the coherence length of edge states in the integer quantum Hall regime, \href{https://doi.org/10.1103/PhysRevLett.100.126802}{Phys. Rev. Lett. \textbf{100}, 126802 (2008)}.

	\bibitem{Probing_2022_Deng}
	P. Deng, C. Eckberg, P. Zhang, G. Qiu, E. Emmanouilidou, G. Yin, S. K. Chong, L. Tai, N. Ni, and K. L. Wang, Probing the mesoscopic size limit of quantum anomalous Hall insulators, \href{https://doi.org/10.1038/s41467-022-31105-w}{Nat. Commun. \textbf{13}, 4246 (2022)}.

	\bibitem{Four_1986_Buttiker}
	M. Büttiker, Four-Terminal Phase-Coherent Conductance, \href{https://doi.org/10.1103/PhysRevLett.57.1761}{Phys. Rev. Lett. \textbf{57}, 1761 (1986)}.

	\bibitem{Dissipation_2024_Liu}
	P.-Y. Liu and Q.-F. Sun, Dissipation and dephasing in quantum Hall interferometers, \href{https://doi.org/10.1103/PhysRevB.110.085411}{Phys. Rev. B \textbf{110}, 085411 (2024)}.

	\bibitem{Landauer_1992_Meir}
	Y. Meir and N. S. Wingreen, Landauer formula for the current through an interacting electron region, \href{https://doi.org/10.1103/PhysRevLett.68.2512}{Phys. Rev. Lett. \textbf{68}, 2512 (1992)}.

	\bibitem{Electronic_1995_Datta}
	S. Datta, Electronic Transport in Mesoscopic Systems (Cambridge University Press, Cambridge, 1995).

	\bibitem{Transport_2022_Zhou}
	H. Zhou, H. Li, D.-H. Xu, C.-Z. Chen, Q.-F. Sun, and X. C. Xie, Transport theory of half-quantized Hall conductance in a semimagnetic topological insulator, \href{https://doi.org/10.1103/PhysRevLett.129.096601}{Phys. Rev. Lett. \textbf{129}, 096601 (2022)}.

	\bibitem{Experimental_2022_Mogi}
	M. Mogi, Y. Okamura, M. Kawamura, R. Yoshimi, K. Yasuda, A. Tsukazaki, K. S. Takahashi, T. Morimoto, N. Nagaosa, M. Kawasaki, Y. Takahashi, and Y. Tokura, Experimental signature of the parity anomaly in a semi-magnetic topological insulator, \href{https://doi.org/10.1038/s41567-021-01490-y}{Nat. Phys. \textbf{18}, 390 (2022)}.

	\bibitem{Dephasing_2023_Fang}
	J.-Y. Fang, A.-M. Guo, and Q.-F. Sun,
	Dephasing effect promotes the appearance of quantized Hall plateaus,
	\href{https://iopscience.iop.org/article/10.1088/1367-2630/acbed2}{New J. Phys.  \textbf{25}, 033001 (2023)}.

	\bibitem{Voltage_1991_McLennan}
	M. J. McLennan, Y. Lee, and S. Datta, Voltage drop in mesoscopic systems: A numerical study using a quantum kinetic equation, \href{https://doi.org/10.1103/PhysRevB.43.13846}{Phys. Rev. B \textbf{43}, 13846 (1991)}.

	\bibitem{Weak_1986_Chakravarty}
	S. Chakravarty and A. Schmid, Weak localization: The quasiclassical theory of electrons in a random potential, \href{https://doi.org/10.1016/0370-1573(86)90027-X}{Phys. Rep. \textbf{140}, 193 (1986)}.

	\bibitem{Phase_1990_Stern}
	A. Stern, Y. Aharonov, and Y. Imry, Phase uncertainty and loss of interference: A general picture, \href{https://doi.org/10.1103/PhysRevA.41.3436}{Phys. Rev. A \textbf{41}, 3436 (1990)}.

	\bibitem{Coupled_1999_Burkard}
	G. Burkard, D. Loss, and D. P. DiVincenzo, Coupled quantum dots as quantum gates, \href{https://doi.org/10.1103/PhysRevB.59.2070}{Phys. Rev. B \textbf{59}, 2070 (1999)}.

	\bibitem{Topological_2013_Bernevig}
	B. A. Bernevig and T. L. Hughes, Topological Insulators and Topological Superconductors (Princeton University Press, Princeton, NJ, 2013).

	\bibitem{Interplay_2025_Wan}
	Y.-H. Wan, P.-Y. Liu, and Q.-F. Sun, Interplay of altermagnetic order and wilson mass in the dirac equation: Helical edge states without time-reversal symmetry, \href{https://doi.org/10.1103/s6pj-495v}{Phys. Rev. B \textbf{112}, 115412 (2025)}.

	\bibitem{Helical_2025_Wan}
	Y.-H. Wan, C.-M. Miao, P.-Y. Liu, and Q.-F. Sun, Helical Fermi arc in altermagnetic Weyl semimetal, \href{https://doi.org/10.1103/bdwz-kmyb}{Phys. Rev. B \textbf{112}, 235411 (2025)}.

	\bibitem{Emergent_2026_Liu}
	P.-Y. Liu, Y.-H. Wan, and Q.-F. Sun, Emergence of net chirality in a two-dimensional Dirac fermion system with altermagnetic mass, \href{https://doi.org/10.1103/k6th-nyt9}{Phys. Rev. B \textbf{113}, L041402 (2026)}.

	\bibitem{Altermagnetism_2025_Wan}
	Y.-H. Wan and Q.-F. Sun, Altermagnetism-induced parity anomaly in weak topological insulators, \href{https://doi.org/10.1103/PhysRevB.111.045407}{Phys. Rev. B \textbf{111}, 045407 (2025)}.

	\bibitem{Half_2022_Zou}
	J.-Y. Zou, B. Fu, H.-W. Wang, Z.-A. Hu, and S.-Q. Shen, Half-quantized Hall effect and power law decay of edge-current distribution, \href{https://doi.org/10.1103/PhysRevB.105.L201106}{Phys. Rev. B \textbf{105}, L201106 (2022)}.
	

	\bibitem{Imaging_2019_Marguerite}
	A. Marguerite, J. Birkbeck, A. Aharon-Steinberg, D. Halbertal, K. Bagani, I. Marcus, Y. Myasoedov, A. K. Geim, D. J. Perello, and E. Zeldov, Imaging work and dissipation in the quantum Hall state in graphene, \href{https://doi.org/10.1038/s41586-019-1704-3}{Nature \textbf{575}, 628 (2019)}.

	\bibitem{Thermal_2021_Fang}
	J.-Y. Fang, N.-X. Yang, Q. Yan, A.-M. Guo, and Q.-F. Sun, Thermal dissipation in the quantum Hall regime in graphene, \href{https://doi.org/10.1103/PhysRevB.104.115411}{Phys. Rev. B \textbf{104}, 115411 (2021)}.

	\bibitem{Current_2017_Kawamura}
	M. Kawamura, R. Yoshimi, A. Tsukazaki, K. S. Takahashi, M. Kawasaki, and Y. Tokura, Current-Driven Instability of the Quantum Anomalous Hall Effect in Ferromagnetic Topological Insulators, \href{https://doi.org/10.1103/PhysRevLett.119.016803}{Phys. Rev. Lett. \textbf{119}, 016803 (2017)}.

\end{thebibliography}
\end{document}